**Impact of Regional Reactions to War on Contemporary Chinese Trade**

Informative title: Impact of War on Chinese Trade

Xuejian Wang

Zhejiang University of Finance and Economics, School of Economics, Hangzhou, China

Email: 180507080003@zufe.edu.cn

Abstract

Different regional reactions to war in 1894 and 1900 can significantly impact Chinese imports in 2001. As international relationship gets tense and China rises, international conflicts could decrease trade. We analyze impact of historic political conflict. We measure regional change of number of people passing imperial exam because of war. War leads to an unsuccessful reform and shocks elites. Elites in different regions have different ideas about modernization, and the change of number of people passing exam is quite different in different regions after war. Regional number of people passing exam increases 1% after war, imports from then empires decrease 2.050% in 2001, and this shows impact of cultural barrier. Manufactured goods can be impacted because brands can be identified easily. Risk aversion of expensive products in conservative regions can increase imports of equipment. Value chains need deep trust, and this decreases imports of foreign company and assembly trade.

# 1. Introduction

The trade war shows huge influence of politics on international trade. Trade happens in real world, and countries have many international conflicts. The situation is more complex than before, because geopolitical factors interact with Chinese regime. We can learn from historic interaction between Chinese regime and international conflicts.

There are severe conflicts involved with Chinese in 19th and 20th century, and the historic reaction of Chinese regions to Western countries can influence Chinese imports from them. This reflects how people view history, and this can influence trade. This shows how cultural bias and political attitude can influence trade in the long term. This also implies how the current conflicts could influence trade in the future. This can explain how different regions have different ways of modernization.

In this paper, the historic regional reactions to war are measured by change of number of people who passed imperial exam because of war (1894-1895). The war is unique because this shook the Chinese elites then, and Chinese empire unexpectedly lost to Japan which has a deeper reform. Deep reform was asked by people who took the exam in Beijing, they sent a letter to emperor and led an unsuccessful short reform later. The war and reform did make the exam less important, because the exam generally focuses on traditional Confucian theory. Importantly, the effect of modernization and reactions to Western countries is different in different regions. In addition, the war in 1900 could deepen the effect.

When a region can modernize from western shock quickly, this region is more likely to trade with Western countries. They could understand Western economic and political ideas, and they could want to develop like Western countries. The effect could be influential even after 100 years. This is because tradition values are passed to next generation. In addition, people with certain values can build up institution, and labor force movement is relatively limited before 2001.

This paper can become part of literature about how international political conflicts influence trade. On the one hand, there are a few results which show no or temporary negative impact on trade (Davis and Meunier, 2011, Du et al., 2017, Fuchs and Klann, 2013, Michaels and Zhi, 2010, Heilmann, 2016, Li and Liu, 2017, Lin et al., 2018, Pandya and Venkatesan, 2015). The results can be explained by liberalism. On the other hand, the impact could be server in other studies (Berger et al., 2013, Blomberg and Hess, 2006, Che et al., 2015, Glick and Taylor, 2010).This can be explained by behaviorism or institutionalism. In addition, China was semi-colony then, so China could get similar impact like colony (Acemoglu et al., 2001, Banerjee and Iyer, 2005, Head et al., 2010, Berthou and Ehrhart, 2017, Bonfatti, 2017).

In this paper, people can know how historic war and reactions impact present international trade. The paper also analyzes different impacts on different types of trade. In addition, we can learn from different impacts on different products and types of firms. These could show how the reactions to war impact trade.

It is the first time that scholars have analyzed how combination of war and modernization impacts contemporary trade in China. The impact on imports is significant for manufactured goods. The effect is significant for foreign firm and assembly trade. This can reflect potential mechanism. These show how historic reactions to international conflicts influence contemporary trade. This brings new evidence on political impact of trade.

Data of exam is the regional increase rate of people passing imperial examination because of wars. Only 4 exams right before the war and 4 exams right after the war are used. This can reflect how regional elites react to Western countries. Western knowledge and culture can be more easily accepted by elites who dropped the exam quickly. The exam increase rate is between -14.3% and 69.2%. The name and region of people passing the exam come from Jiang (2007). The province boarder changes a little from Qing Dynasty (1644-1911)(Tanet al. 1982), and we adjust this effect.

The change of imperial exam could impact trade with Western countries, and Western countries are represented by the major empires at that time. Empires analyzed in this paper include British Empire, France, Germany, Italy, Japan and United States. They are 6 of 8 countries from Eight-Nation Alliance which fought with China in 1990. Russia and Austro-Hungary are dropped for historical reasons.

In our model, regional reactions to war can impact international trade by cultural cost, risk-aversion and value-chain effects. Additional cultural cost could mean that conservative regions trust empires less, and this will increase relative price and decrease trade. This shows impact of cultural bias and trust on trade (Che et al., 2015, Guiso et al., 2009). Imports from empires can be impacted more than exports, because Chinese are impacted more by the wars, and people tend to be more biased toward imports than exports (Mansfield et al., 2016, Wall and Heslop, 1986).

Risk-aversion effect means that risk aversion of expensive products will increase imports of conservative regions from empires. This is because when people try to buy expensive products, they show more risk aversion (Kahneman and Tversky, 1979, Hans, 1980, Kachelmeier and Shehata, 1992). The limited information could also increase risk aversion (Brown et al., 1988). They could get more utility from famous expensive products. Former empires tend to have large enterprises and famous products.

Value-chain effect means that additional cultural cost can decrease imports of conservative regions further because of value-chain. China imports inputs from other countries and assembles them (Tung and Wan Jr, 2013, Yu and Luo, 2018). Value chains need more trust and cooperation, and distrust can decrease the number of enterprises which locate in conservative regions. This can decrease the additional demand of imports from value chains.

In empirical model, we only use the trade data in 2001 from custom, because China joined WTO in 2001. The data we use can eliminate effect of WTO, and trade value is relatively large in 2001.

We can analyze the impact of reactions from war by gravity model, which is estimated by difference in differences (DD) and Poisson pseudo maximum likelihood (PPML). DD makes difference of trade between empire and other countries, as well as difference among different regions. The first difference cuts other factors related with regions, and the second difference cuts the other factors related with countries. Meanwhile, country and region fixed effects are controlled. In addition, we can analyze mechanism behind the process by difference in difference in differences (DDD).

Baseline results show that when people passing the exam increase 1% in one region, total trade decreases 1.027% and imports decrease 2.050% from empires. This reflects the impact of cultural bias (Che et al., 2015, Guiso et al., 2009). Additional cultural cost effect can lower imports of conservative regions from empires.

Further analysis shows that conservative regions import more manufactured goods from empires, and this reflects strong cultural cost effect because of brand. Conservative regions can increase more equipment from empires, and this shows risk aversion of expensive equipment. Not surprisingly, imports of foreign enterprises from empires decrease more, and this shows effect of cultural cost or value chains. State-owned enterprises of conservative regions can import more from former empires, and this shows risk aversion of state-owned enterprises. Lastly, assembly trade of conservative regions from empires can decrease more, processing with inputs does not decrease significantly, and this shows value-chain effect of assembly trade.

The results are robust for regions which are not directly involved in war of 1900. We get similar results from Chinese trade with all developed countries. The results without Japan are also similar. However, when we compare Sino-Japanese trade with other trade partners, we get significant negative results of imports and exports from Japan.

For future research, this paper shows political and cultural impact from Chinese evidence, and we also need evidence from other developing countries. The mechanism behind trade, politics and modernization needs to be analyzed further.

Session 2 is historic background and theory. Data is described in session 3. Session 4 shows empirical model. Baseline results and trade structure are in session 5 and 6. Session 7 reports robust test.

## 2. Historical Background and Theory

Firstly, we show historical background. In addition, we combine history with cultural cost, risk aversion and value chains, and we get the model and propositions.

### 2.1 Historical Background

The impact of war is measured by increase rate of people passing imperial examination. The exam is essential way to hire new officials from 587 D.C. to 1905, but it is quite conservative in 19$^{th}$ century. The exam generally focuses on Confucianism. The theory is not enough to save dying dynasty of China in 19$^{th}$ century. The empire needs new ideas, institutions and weapons. Some elites understood this better than others.

The modernization movement changed suddenly for unexpected failure of Sino-Japanese war (1894-

1895), and this brings up a chance to test regional difference in reactions of modernization. Modernization has started during Self-Strengthening Movement (1861 – 1895), but the war in 1894 shows that the movement fails. The failure is awkward and unexpected for Chinese, because some elites thought that they did start industrialization and China was still dominant power in Eastern Asia. The war makes some elites want more change, but they disagree with how to change. War in 1900 could deepen the effect. The different reactions in different regions are shown by exam, and this creates exogenous variable for this paper.

Although there are many wars in China in 19th and 20th century, this war changes the people who takes the exam deeply. When people heard the news of peace deal of war in 1895, people who took the national exam in Beijing organized to ask emperor to reform. The empire did an unsuccessful reform led by people who took the exam. After reform failed, the reformers were either killed or left China, and the exam ended in 1905.

The war and modernization impact Chinese a lot, and this can make different regions trade differently with so-called empires (global powers). The empires only include 6 countries of major global powers at that time, and they joined the war between China and Eight-Nation Alliance from 1899 to 1900. All 6 nations include United States, British Empire, Germany, France, Italy and Japan. However, Russia is excluded because of Soviet Union, and Austro-Hungary is excluded because of collapse.

Empires could trade with some traditional regions less, because cultural bias from the war decrease trade. The better understanding of modernization could make some regions industrialize quickly. They could join global value chains and get investment more easily. Meanwhile, conservative regions have less information, and they could buy more expensive products from empires to lower risks.

Some may doubt why the war 100 years ago can still impact trade, this can happen for family tradition, limited labor movement and institution. Firstly, the family tradition in China makes people have special feelings about family and hometown. Traditional values still tend to make people stay at home and share the basic values with family members. In addition, the market economy in China is not so mature in 2001, and we use trade data in 2001. China was not member of WTO then, and China suffered from several wars and planned economy. These make people hard to leave their home town. Lastly, the basic ideas of local elites could impact institution, and the institution can impact youngsters.

## 2.2 Theory

### 2.2.1 Cultural Trade Cost

A typical region can get utility from imports, and we get the utility function and budget line.

$$\max_{M^*,\ M} U = M^{*\alpha} M^{1-\alpha} \quad (1)$$

$$s.t. (p^* + c)M^* + pM = Y \quad (0 < \alpha < 1) \quad (2)$$

M* means imports of a Chinese region from global powers (empires), M stands for imports from other countries, and $\alpha$ stands for elasticity. The average price of imports from global powers is p*, average price of other imports is p, and Y stands for budget of imports. We use c stands for additional cultural cost of imports from global powers.

There is additional cost (c) of imports from global powers, because cultural biases impact trade (Che et al., 2015, Guiso et al., 2009). Historical conflicts can hurt people's trust, and this can lower trade. People pass the values to youngsters, and institution also makes youngsters remember history and tradition. Therefore, war can impact trade in the long term.

In addition, we assume that imports are influenced more than exports. This is because people tend to be more biased toward imports than exports (Mansfield et al., 2016, Wall and Heslop, 1986). In addition, China is invaded by other empires, so some conservative Chinese could be hurt more and import less from empires. Then we get first order condition.

$$\alpha M^{*(\alpha-1)} M^{1-\alpha} + \lambda(p^* + c) = 0 \quad (3)$$

$$(1-\alpha) M^{-\alpha} M^{*\alpha} + \lambda p = 0 \quad (4)$$

$$(c + p^*) M^* + pM = Y \quad (5)$$

Then we get demand function of imports from global powers and other countries.

$$M^* = \frac{\alpha Y}{P^* + c} \quad (6)$$

$$M = \frac{(1-\alpha) Y}{P} \quad (7)$$

The additional trade cost can decrease the imports from empires. When regional number of people taking imperial exam increases more, the region has higher cultural trade costs with empires. The higher cultural costs can increase relative price of imports, and this can lower the imports from empires.

Proposition 1 (cultural cost effect): If regional number of people taking imperial exam increases more, the region will have higher cultural trade costs, and the region will import less from empires.

### 2.2.2 Information and Risk Aversion

Different products could have different effects. According to prospect theory of behavioral economics, people tend to show risk aversion, because they get more impact by losses than gains (Kahneman and Tversky, 1979). People show more risk aversion when the payoffs are higher (Hans, 1980, Kachelmeier and Shehata, 1992). In this paper, we get the utility function of different products.

$$\max_{M_N^*, M_E^*, M_N, M_E} M_N^{*\alpha(1-\beta)} M_E^{*\alpha\beta} M_N^{(1-\alpha)(1-\beta)} M_E^{(1-\alpha)\beta} \quad (8)$$

$$s.t. (c + p_N^*) M_N^* + (c + p_E^*) M_E^* + p_N M_N + (p_E + d) M_E = 0 \quad (0 < \alpha, \beta < 1, d > 0) \quad (9)$$

In the equation, $M_N^*$ stands for imports of normal products from empires, $M_E^*$ stands for imports of expensive products from empires, $M_N$ stands for imports of normal products from other countries, $M_E$ stands for imports of expensive products from other countries. We use $p$ to show different prices of different products. In addition, conservative regions know less about foreign countries, and former empires have many famous and large enterprises. Therefore, we use $d$ to measure risks of buying expensive products from other countries. We use $\beta$ to measure utility and show risk aversion of expensive products.

Because people show more risk aversion when the payoffs are higher (Hans, 1980, Kachelmeier and Shehata, 1992), we use $\beta$ and $d$ to show risk aversion when people buy expensive products. We use $\beta$ to change the elasticity of expensive product. When $\beta$ is higher, the utility of expensive products is higher. The high utility of expensive products shows expensive products are important, and $d$ show the risk of buying expensive products from other countries. In addition, conservative regions have less information, and they have less chance to diversify (Brown et al., 1988). Conservative regions can be more sensitive to risks, and $\beta$ and $d$ are higher for conservative regions. Then we get the first order condition.

$$M_N^{*\alpha(1-\beta)-1} M_E^{*\alpha\beta} M_N^{(1-\alpha)(1-\beta)} M_E^{(1-\alpha)\beta} + \lambda(c + p_N^*) = 0 \quad (10)$$

$$M_N^{*\alpha(1-\beta)} M_E^{*\alpha\beta-1} M_N^{(1-\alpha)(1-\beta)} M_E^{(1-\alpha)\beta} + \lambda(c + p_E^*) = 0 \quad (11)$$

$$M_N^{*\alpha(1-\beta)} M_E^{*\alpha\beta} M_N^{(1-\alpha)(1-\beta)-1} M_E^{(1-\alpha)\beta} + \lambda p_N = 0 \quad (12)$$

$$M_N^{*\alpha(1-\beta)} M_E^{*\alpha\beta} M_N^{(1-\alpha)(1-\beta)} M_E^{(1-\alpha)\beta-1} + \lambda(p_E + d) = 0 \quad (13)$$

In addition, we get demand functions.

$$M_E^* = \frac{\alpha\beta Y}{c + p_E^*} \quad (14)$$

$$M_N^* = \frac{\alpha(1-\beta)Y}{c + p_N^*} \quad (15)$$

$$M_E = \frac{(1-\alpha)\beta Y}{p_E + d} \quad (16)$$

$$M_N = \frac{(1-\alpha)(1-\beta)Y}{p_N} \quad (17)$$

If the effect of risk aversion is high, and risk of buying expensive products ($\beta, d$) is high, conservative regions could import more expensive products ($M_E^*$) from empires. This is because if a region is really conservative, the region will know less about foreign countries. They could be worried about risk of expensive products from other countries. In addition, the region shows higher risk version because of high price and little information. The effects could be larger than effect of cultural cost. For imports from empires, we need to compare the cultural cost and risk-aversion effects.

Proposition 2 (risk-aversion effect): If conservative regions have limited information of foreign products, and the regions show risk aversion of expensive products (higher $\beta, d$), the regions will import more expensive products from empires.

### 2.2.3 Global value chains and product

In global value chains, trust can deeply impact cooperation of different countries. China is the world factory, China imports inputs from other countries, and assembles them to produce final product (Yu and Luo, 2018, Tung and Wan Jr, 2013). Then we get utility function and budget line to show impact of value chains.

$$\max_{M_N^*, M_E^*, M_N, M_E} (\gamma M_N^*)^{\alpha(1-\beta)} (\gamma M_E^*)^{\alpha\beta} M_N^{(1-\alpha)(1-\beta)} M_E^{(1-\alpha)\beta} \quad (18)$$

$$s.t. (c + p_N^*)\gamma M_N^* + (c + p_E^*)\gamma M_E^* + p_N M_N + (p_E + d)M_E = 0 \quad (0 < \alpha, \beta < 1, d > 0, \gamma > 1) \quad (19)$$

We use $\gamma$ to measure the additional cultural cost of certain trade type from empires because of value chains. Because the value chains need deep collaboration of China and other countries, trust is more important. Distrust can hurt collaboration, decrease the need of import of inputs, lower the utility of imports, and increase costs of collaboration. Then we get demand functions.

$$M_E^* = \frac{1}{\gamma} \frac{\alpha\beta Y}{c + p_E^*} \quad (20)$$

$$M_N^* = \frac{1}{\gamma} \frac{\alpha(1-\beta)Y}{c + p_N^*} \quad (21)$$

$$M_E = \frac{(1-\alpha)\beta Y}{p_E + d} \quad (22)$$

$$M_N = \frac{(1-\alpha)(1-\beta)Y}{p_N} \quad (23)$$

If $\gamma$ increases, the region will import less from empires when the income is constant. This means that there are less utility and more cost from imports, and this increases relative prices and decreases the demand of inputs.

Proposition 3 (value-chain effect): If certain type of trade relies on collaboration (higher $\gamma$), this will decrease the imports of conservative regions from empires. This means additional cultural costs lower the demand of inputs further because of value chains.

All in all, to analyze impact on trade, we need to compare three effects above. For example, when we analyze expensive imports from empires, we can find negative cultural cost effect (proposition 1), positive risk-aversion effect (proposition 2) and negative value-chain effect (proposition 3).

## 3. Data

In this part, we discuss the general information of data firstly, and we discuss huge regional difference later.

### 3.1 basic information

The basic attitudes towards Western countries can be reflected in change of numbers of people passing the exam. The regional information of people who passed the exam is from Jiang (2007). The measurement focuses person who passed the exam right before war and right after war. China lost the war (1894-1895) mostly in 1894, and the war ended in the spring of 1895. Therefore, 4 exams before the war (1889-1894) and 4 exams after the war (1895-1904) are included. This is because there are only 4 exams after the war. We add 4 exams before the war to compare. We can adjust the change of boarder of province from Qing Dynasty (Tanet al. 1982).

The war and reform could make some people drop the exam. Wars can reflect that traditional ideas are not enough to make China great again, but some people have better understanding of this idea. However, imperial examination is a traditional exam of becoming officials, so benefits and traditional values make people hard to quit the exam.

There are different opinions about modernization, and reactions to wars reflect elites' ideas. Because elites can pass the opinions to young elites directly, and they can use these ideas to build up institution. The institution can make the general public hold the idea. The family values and relatively limited labor force movement make ideas more powerful.

We use trade data from China custom in 2001. This is because China joined WTO in 2001. There is a lot of trade between China and world in 2001, and we can avoid the effect of WTO. Trade with Hong Kong, Macau and tax-haven countries or areas are excluded.

**3.2 Regional difference**

Certainly, elites in different regions have different ideas about how to develop. Open-minded officials are usually more open-mined about modernization, while conservatives doubt this. Coastal and inland provinces could have different ideas, but each province also has its own characteristics.

After the war, the numbers of people from some province passing the exam increase dramatically,

while others decrease relatively (table 1). While only the number in Shanghai decreased, the increase rates are quite different in different regions. Some provinces can be above 50%, while others can be around 0.

Table 1 Regional change of people passing imperial exam because of war

| Region | Before war | After war | Increase rate |
| --- | --- | --- | --- |
| ANHUI | 60 | 69 | 0.150 |
| BEIJING | 23 | 25 | 0.087 |
| CHONGQING | 13 | 15 | 0.154 |
| FUJIAN | 68 | 94 | 0.382 |
| GANSU | 32 | 38 | 0.188 |
| GUANGDONG | 53 | 72 | 0.358 |
| GUANGXI | 46 | 57 | 0.239 |
| GUIZHOU | 40 | 47 | 0.175 |
| HEBEI | 65 | 110 | 0.692 |
| HENAN | 63 | 73 | 0.159 |
| HUBEI | 48 | 67 | 0.396 |
| HUNAN | 45 | 61 | 0.356 |
| JIANGSU | 78 | 86 | 0.103 |
| JIANGXI | 75 | 89 | 0.187 |
| JILIN | 3 | 3 | 0 |
| LIAONING | 12 | 12 | 0 |
| SHAANXI | 41 | 60 | 0.463 |
| SHANDONG | 75 | 94 | 0.253 |
| SHANGHAI | 14 | 12 | -0.143 |
| SHANXI | 32 | 45 | 0.406 |
| SICHUAN | 37 | 47 | 0.270 |
| TIANJIN | 27 | 28 | 0.037 |
| YUNNAN | 56 | 73 | 0.304 |
| ZHEJIANG | 78 | 100 | 0.282 |

The war is Sino-Japanese War in 1894-1895.
Table includes 4 exams right before war and 4 exams right after war.

Although eastern developed provinces generally have lower increase rate, this does not hold in every case. Some eastern provinces are interested in the exam (e.g. Fujian and Hebei), and some inland provinces dislike the exam (e.g. Anhui). This reflects the interest about modernization beyond geography. Regions which prefer the exam can be quite different. The regions could be coastal or inland. However,

Northern and Central provinces are more likely to like the exam. In contrast, people in coast areas or big cities could drop the exam.

Apparently, there is missing data of some provinces. This is because some provinces are quite different from now. There are many minorities in some regions, and they do not take the exam as much as Han Chinese (majority). Some provinces are less populated like now, and the numbers of their people passing the exam are 0 or nearly 0. People from Banner system (military system of Qing) are dropped, because the region information is missing.

The regional increase in exam is negatively correlated with share of trade with developed countries in 2001 (Figure 1). Although the points are quite scattered, both exports and imports show negative relation. The slope of exports is bigger than the slope of imports. All in all, we need more rigorous analysis.

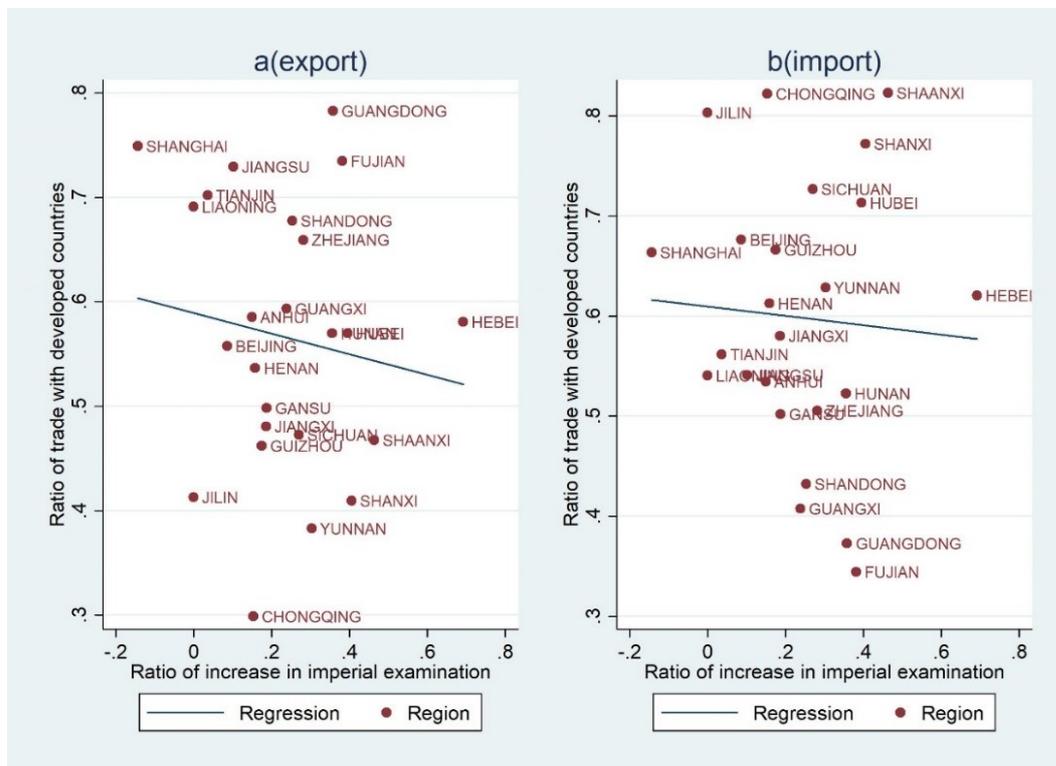

**Figure 1** Correlation between ratio of trade with developed countries and increase rate of exam

## 4. Empirical model

We use difference in differences (DD) and Poisson pseudo maximum likelihood (PPML). More detailed cause of trade is analyzed by difference in difference in differences (DDD). The model is based on another model (Che et al., 2015). In addition, we control region and country fixed effects. We use clustered standard errors of region. To get unbiased results, we use the estimation equation:

$$y_{rk} = \beta_0 + \gamma * increase_r * empire + FE_{region} + FE_{country} + \mu_{rk} \quad (25)$$

In the equation, *y* stands for trade value. We use logarithm of value of trade in OLS results, while we directly use value of trade in PPML. *Increase* stands for regional increase rate of people passing imperial exam after the war. The *empire* stands for 6 major global powers at that time, including British Empire, France, Germany, Italy, Japan and United States. However, Russia and Austro-Hungary are excluded because of historical reasons. The interaction term between increase and empire shows the impact.

Region fixed effect can control stable variables in certain region. This can control distance to Beijing, economic factors, technical trend and other cultural factors. Similarly, country fixed effect can control stable variables in a certain country. This controls distance to China, economic factors, technical trend and cultural factors.

After controlling fixed effect, we can get unbiased results from DD. When we get difference of trade between empire and other countries, we can eliminate potential endogenous regional factors. When we get difference of trade among different Chinese regions, we can eliminate potential endogenous factors related with countries.

DDD method is based on DD, and this can divide trade into more detailed categories. When the effect is significant for certain types of trade, we can know how and why the effect happens. In equation, we control paired 2-dimensional fixed effects between region, country or new category. In addition, we add 3 paired interaction terms between categories, region or country. We also add interaction term among categories, region and countries, and the coefficient show the impact.

We use PPML to estimate the results, we can solve problems of too many zero values, and this can also avoid inconsistent estimates in the presence of heteroskedasticity (Silva and Tenreyro, 2006).

## 5. Baseline

The baseline results show that increase in exam can decrease trade with empire significantly (table 2). In OLS regression, coefficients are not significant, while coefficients of total value of trade and imports are both significant at 1% significant level in PPML. Considering the bias of zero trade value and inconsistent estimates in the presence of heteroskedasticity, results in PPML are more convincing. The coefficients in PPML are significant both economically and statistically. We use PPML only in later results.

Table 2 Influence of increase in imperial examination on trade for empire

|  | (1) OLS all | (2) OLS exports | (3) OLS imports | (4) PPML all | (5) PPML exports | (6) PPML imports |
|---|---|---|---|---|---|---|
| Increase* empire | -0.523 | -0.378 | -0.615 | -1.027*** | -0.124 | -2.050*** |
|  | (0.463) | (0.425) | (0.560) | (0.103) | (0.558) | (0.300) |
| Region FE | YES | YES | YES | YES | YES | YES |
| Country FE | YES | YES | YES | YES | YES | YES |
| Number of regions | 24 | 24 | 24 | 24 | 24 | 24 |
| Number of countries | 189 | 189 | 189 | 189 | 189 | 189 |
| n | 3696 | 3625 | 2081 | 4536 | 4536 | 4536 |

Country and region fixed effects are controlled and clustered standard errors of region are in parentheses. Independent variable of OLS is ln(trade value), while independent variable of PPML is value. Increase rate is from 4 exams right before war and 4 exams right after war.    * $p < 0.1$, ** $p < 0.05$, *** $p < 0.01$.

This shows that if the number of people passing the exam increase 1%, regional total trade will decrease 1.027% and imports will decrease 2.050% from empires. Considering large variance of increase rate (from -14.3% to 69.2%), the effect is relatively important.

This could show that regions with lower increase rate understand empires better, and this can lower the cultural trade cost (proposition 1). This means that deep cultural barrier can decrease international trade (Che et al., 2015, Guiso et al., 2009). The cultural trade cost can make people in conservative regions trust empires less, and it's hard to make deals. The cultural cost could increase cost of imports from empires, and this lowers the imports from empires.

In addition, the results are significant for imports, while the coefficients of exports are not significant. This shows that people with better image of empires can import more, while this does not impact exports from empires. This is because people tend to have more biases against imports than exports (Mansfield et al., 2016, Wall and Heslop, 1986). Conservative regions have more biases of imports of empires. This also reflects that Chinese are victims of the wars, and empires are less influenced by the wars. In addition, we can get similar ideas from coefficient of total trade value. This is because the coefficient of imports is larger than coefficient of total trade value. In addition, we need more detailed data of trade to test proposition 2 and 3.

**6. Trade structure**

More detailed analysis of trade data can show how reactions to the war impact trade. In this part, we analyze special impact from product type, enterprise type and trade type. This can test the theory above.

**6.1 Product type**

Table 3 Different product and influence of increase in imperial examination on imports (PPML)

| SITC0 | SITC1 | SITC2 | SITC3 | SITC4 |
|---|---|---|---|---|
| 1.208 | -3.639 | 0.876 | 0.169 | 2.327 |
| (1.358) | (3.274) | (0.817) | (0.960) | (2.974) |

| SITC5 | SITC6 | SITC7 | SITC8 | SITC9 |
|---|---|---|---|---|
| -0.283 | -1.102** | 0.480* | 0.220 | 32.22*** |
| (0.273) | (0.487) | (0.261) | (0.445) | (3.002) |

Sample size is 45360. Three 2-dimensional fixed effects between country, region or product are controlled. Clustered standard errors of region are in parentheses. Independent variable of PPML is trade value. Increase rate is from 4 exams right before war and 4 exams right after war. SITC0-9 denotes: food, live animals; beverages and tobacco; crude materials, inedible, except fuels; mineral fuels, lubricants and related materials; animal and vegetable oils, fats and waxes; chemicals and related products; manufactured goods classified chiefly by material; machinery and transport equipment; miscellaneous manufactured articles; not classified elsewhere.
* $p < 0.1$, ** $p < 0.05$, *** $p < 0.01$.

In this part, imports are divided according to SITC classification. Similar analysis and classification can be found in other research (Lin et al., 2017). There are total 10 types of products. Only two results are significant (table 3). SITC6 stands for manufactured goods, and the coefficient is significantly negative. Meanwhile, machinery and transport equipment (SITC7) are significantly positive, but the effect is relatively small.

We use DDD to get results in this part. Specifically, we add another group classification (i.e. belong to SITC5 or not), and we add three paired intersection terms between these three variables. We also add intersection term of new variable, increase rate and empire dummy, and the coefficient shows the results of DDD. We control three paired 2-dimensional fixed effects between region, country or product.

Manufactured goods (SITC6) decrease 1.102% when the number of people passing exam increases 1% (table 3), and this shows effect of cultural cost (proposition 1). People can easily find out where manufactured goods come from, because the manufactured goods have brands. When they notice brands, they can know where the products come from, and have certain biased opinions about products. In addition, manufactured goods are not especially expensive, so the effect of risk aversion is not really big (proposition 2). People do not need to buy famous brand from empires, because the products are relatively cheap and people can take risks.

The abnormal positive coefficient (SITC7) could reflect that risk aversion and limited information lead to increase in imports of equipment from empires (proposition 2). While we can still find the effect of cultural cost (proposition 1), the effect could be much smaller than risk-aversion effect (proposition 2). The risk aversion is important when people try to buy expensive products, and machinery and transport equipment belong to these products. When they buy these expensive products, they show risk aversion. In addition, they have limited information and could only know famous products. Expensive famous products are usually from former empires.

In addition, significant results for SITC9 are meaningless, because SITC9 stands for goods which are not classified. The value of goods in SITC9 is quite low. Insignificant coefficients of other product show that the cultural cost effect could counteract risk-aversion effect, or two effects are not really large.

**6.2 firm type**

We analyze three types of firm: state-owned enterprise, private enterprise and foreign enterprise. Foreign enterprise is enterprise which is owned by foreign partly or completely, and enterprise can be partnership or corporation. The method of analysis here is similar with analysis of product type.

Coefficients of state-owned and foreign enterprise are significant, but only coefficient of foreign is negative (table 4). The coefficient of foreign enterprise is a little larger than state-owned enterprise. Both regressions are significant at 5%. The coefficient for private enterprise is not significant.

Table 4 Different firm type and influence of increase in imperial examination on imports (PPML)

|  | (1) SOE | (2) FOREIGN | (3) PRIVATE |
|---|---|---|---|
| intersection | 0.806** | -0.874** | 0.0969 |
|  | (0.340) | (0.401) | (1.715) |
| 2-Dimensional FE | YES | YES | YES |
| Number of regions | 24 | 24 | 24 |
| Number of countries | 189 | 189 | 189 |
| N | 9072 | 9072 | 9072 |

Three 2-dimensional fixed effects between country, region or firm type are controlled. Clustered standard errors of region are in parentheses. Independent variable of PPML is trade value. Increase rate is from 4 exams right before war and 4 exams right after war.
* $p < 0.1$, ** $p < 0.05$, *** $p < 0.01$.

The negative coefficient of foreign enterprise shows the cultural cost effect (proposition 1) and value-chain effect (proposition 3). Many foreign enterprises are part of global value chains, and they import intermediate goods from other countries. The enterprises from empires are less likely to locate in conservative regions because of distrust, so conservative regions import less from home country of foreign enterprises. In addition, the institution and workers of conservative regions can be more conservative than other regions, and this can also decrease the imports from empires. The effect of risk aversion could be very small, because foreign enterprises usually have more information of foreign country to lower risks.

The abnormal positive results of state-owned enterprises could show that state-owned enterprises show risk aversion, and they import more from other foreign large enterprises (proposition 2). Managers of state-owned enterprise show risk aversion, because they do not have incentive to take risks. They could also buy more expensive products and this will make managers lower risks, because state-owned enterprises are generally larger than others in China. In addition, larger enterprises in conservative regions know less about Western countries, they tend to trade with famous large enterprises to lower risks. The empires own more famous large firms than other countries. The significant coefficient shows that the risk-aversion effect is larger than cultural cost effect and value-chain effect.

**6.3 Trade type**

Processing trade is key to the impact of war on trade. Model here is with similar models in session 6.1 and 6.2. We analyze trade type including processing with inputs, assembly and equipment. Similar analysis and classification can be found in another research (Yu, 2015).

Processing trade can be divided into processing with input, assembly, equipment and others. Processing with input is one type of processing trade, and domestic producer buy material and sell products abroad by themselves. In contrast, assembly means that domestic producer only gets assembly fees, and the use of material, equipment, design and selling are controlled by foreign firm. Equipment means that equipment imported, and the equipment could be used for processing trade or investment.

Assembly trade significantly decreases 1.385% when regional exam increases 1% (Table 5). However, the coefficient for equipment is significantly positive, but the coefficient is smaller.

Table 5 Different trade type and influence of increase in imperial examination on imports (PPML)

|  | (1) Processing with inputs | (2) Assembly | (3) Equipment |
|---|---|---|---|
| Intersection | -0.182 | -1.385*** | 1.019* |
|  | (0.124) | (0.362) | (0.562) |
| 2-Dimensional FE | YES | YES | YES |
| Number of regions | 24 | 24 | 24 |
| Number of countries | 189 | 189 | 189 |
| n | 9072 | 9072 | 9072 |

Three 2-Dimensional fixed effects between country, region or trade type are controlled. Clustered standard errors of region are in parentheses. Independent variable of PPML is trade value. Increase rate is from 4 exams right before war and 4 exams right after war.
* $p < 0.1$, ** $p < 0.05$, *** $p < 0.01$.

The significant coefficient of assembly trade shows that value-chain effect decreases the imports of conservative regions from empires (proposition 3). This is because domestic producer only gets assembly fees and foreign firms control other process of assembly trade, so assembly trade needs more trust. Conservative domestic people are unwilling to make foreign corporation control the whole process, and this will lower the imports of assembly trade. In addition, the cooperation and information of foreign partners can lower risks, and this will also decrease imports of conservative regions by lowering the risk-aversion effect (proposition 2).

In contrast, the coefficient of processing with inputs is insignificant, because people can control most of the production process when they are under processing with inputs. Processing with inputs does not need additional trust, so value-chain effect can be smaller than assembly trade. They do not have additional information from foreign partners, and this may increase trade by increasing risk-aversion effect. Cultural cost effect may decrease imports of processing with inputs. Three effects could make the coefficients insignificant.

The abnormal positive sign of equipment could reflect that risk-aversion effect causes more imports of equipment in conservative regions (proposition 2). These regions have to lower the risks because the equipment is so expensive. They could have limited information about equipment and Western countries. Conservative regions want to buy famous equipment. The famous equipment is more likely to come from former empires. While cultural cost and value chain effects could lower the imports, the effects could be smaller than risk-aversion effect.

**7. Robustness**

The results are quite robust in different sets of regression. The first test excludes some special regions in China, second test analyzes influences on all developed countries, and last test analyzes trade of Japan.

Table 6 Main region and influence of increase in imperial examination on trade for empire (PPML)

|  | (1) PPML all | (2) PPML exports | (3) PPML imports |
|---|---|---|---|
| Increase* empire | -1.098*** | -0.151 | -2.083*** |
|  | (0.0671) | (0.536) | (0.180) |
| Region FE | YES | YES | YES |
| Country FE | YES | YES | YES |
| Number of regions | 19 | 19 | 19 |
| Number of countries | 189 | 189 | 189 |
| N | 3591 | 3591 | 3591 |

Country and region fixed effects are controlled and clustered standard errors of region are in parentheses.
Increase rate is from 4 exams right before war and 4 exams right after war.
Main region means regions which are not directly involved in war in 1900.
* $p < 0.1$, ** $p < 0.05$, *** $p < 0.01$.

The first test shows that results excluding regions directly involved in war of 1900 are not much different from baseline results (table 6). The regions directly involved with war in 1900 include Beijing, Tianjin, Hebei, Liaoning and Jilin. While the absolute value of coefficients in this part is a little bigger than baseline results, the difference is quite small.

Besides test in regions, the test of nation does show robustness. We change the empire term (baseline) into term of developed countries. The coefficients with developed countries are quite similar with the coefficients in baseline results (table 7). Different from baseline results, the trade partners here are generally developed country. This includes high-income countries in Western, Northern, Southern Europe, North America, Oceania and Japan.

Table 7 All developed countries and influence of increase in imperial examination on trade (PPML)

|  | (1) PPML all | (2) PPML exports | (3) PPML imports |
|---|---|---|---|
| Increase* developed | -1.171*** | -0.0367 | -2.293*** |
|  | (0.149) | (0.671) | (0.504) |
| Region FE | YES | YES | YES |
| Country FE | YES | YES | YES |
| Number of regions | 24 | 24 | 24 |
| Number of countries | 189 | 189 | 189 |
| n | 4536 | 4536 | 4536 |

Country and region fixed effects are controlled and clustered standard errors of region are in parentheses.
Increase rate is from 4 exams right before war and 4 exams right after war.
Developed countries mean traditional high-income countries in Western, Northern, southern Europe, North America, Oceania and Japan
* $p < 0.1$, ** $p < 0.05$, *** $p < 0.01$.

Table 8 Japan and influence of increase in imperial examination on trade (PPML)

|  | (1) Without Japan all | (2) Without Japan exports | (3) Without Japan imports | (4) Only Japan all | (5) Only Japan exports | (6) Only Japan imports |
|---|---|---|---|---|---|---|
| Increase* empire | -0.647** | 0.658 | -2.325*** |  |  |  |
|  | (0.291) | (0.879) | (0.511) |  |  |  |
| Increase* Japan |  |  |  | -1.314*** | -1.678*** | -1.020*** |
|  |  |  |  | (0.349) | (0.593) | (0.206) |
| Region FE | YES | YES | YES | YES | YES | YES |
| Country FE | YES | YES | YES | YES | YES | YES |
| Number of regions | 24 | 24 | 24 | 24 | 24 | 24 |
| Number of countries | 188 | 188 | 188 | 189 | 189 | 189 |
| N | 4512 | 4512 | 4512 | 4536 | 4536 | 4536 |

Country and region fixed effects are controlled and clustered standard errors of region are in parentheses.
Increase rate is from 4 exams right before war and 4 exams right after war.
* $p < 0.1$, ** $p < 0.05$, *** $p < 0.01$.

Last test of Japan shows a little change of value of coefficients (table 8). The first three results show regression without data of Japan, the absolute value of total trade is a little smaller, but the result for imports is a little bigger.

In addition, we change the term of empire (baseline results) into term of Japan in the last three results. The coefficients for Sino-Japanese trade are all significant. The coefficient of total trade is a little bigger, the coefficient of imports is smaller. Unexpectedly, the negative coefficient of exports is quite significant. Historical tension between China and Japan could bring stronger cultural cost on trade. We need more detailed researches on political factors of trade in China, Japan and other parts of the world.

## 8. Conclusion

If regional increase rate of imperial exam is relatively lower because of war, the region can significantly import more from then global powers in 2001. Further analysis shows that cultural cost and effect of value chain can decrease imports, but risk aversion can increase imports of equipment. We need more historic evidence in these areas.